\documentclass[apj]{emulateapj}
\def\gtorder{\mathrel{\raise.3ex\hbox{$>$}\mkern-14mu
             \lower0.6ex\hbox{$\sim$}}}
\def\ltorder{\mathrel{\raise.3ex\hbox{$<$}\mkern-14mu
             \lower0.6ex\hbox{$\sim$}}}

\usepackage{colordvi}
\usepackage[usenames,dvipsnames]{xcolor}

\slugcomment{Draft of \today}

\shorttitle{SNHunt\,275 (PTF\,13efv)}

\shortauthors{Ofek et al.}

\begin{document}

\title{PTF\,13efv -- An outburst 500 days prior to the SNHunt\,275 explosion and its radiative efficiency}
\author{E.~O.~Ofek\altaffilmark{1},
S.~B.~Cenko\altaffilmark{2},
N.~J.~Shaviv\altaffilmark{3,4},
G.~Duggan\altaffilmark{5},
N.-L.~Strotjohann\altaffilmark{1},
A.~Rubin\altaffilmark{1},
S.~R.~Kulkarni\altaffilmark{5},
A.~Gal-Yam\altaffilmark{1},
M.~Sullivan\altaffilmark{6},
Y.~Cao\altaffilmark{5},
P.~E.~Nugent\altaffilmark{7,8},
M.~M.~Kasliwal\altaffilmark{9},
J.~Sollerman\altaffilmark{10},
C.~Fransson\altaffilmark{10},
A.~V.~Filippenko\altaffilmark{8},
D.~A.~Perley\altaffilmark{11},
O.~Yaron\altaffilmark{1}, and
R.~Laher\altaffilmark{12}}

\altaffiltext{1}{Benoziyo Center for Astrophysics and the Helen Kimmel center for planetary science, Weizmann Institute of Science, 76100 Rehovot, Israel.}
\altaffiltext{2}{Astrophysics Science Division, NASA Goddard Space Flight Center, Mail Code 661, Greenbelt, MD, 20771, USA}
\altaffiltext{3}{Racah Institute of Physics, The Hebrew University, 91904 Jerusalem, Israel}
\altaffiltext{4}{School of Natural Sciences, Institute for Advanced Study, Princeton NJ 08540, USA}
\altaffiltext{5}{Cahill Center for Astronomy and Astrophysics, California Institute of Technology, Pasadena, CA 91125, USA}
\altaffiltext{6}{School of Physics and Astronomy, University of Southampton, Southampton SO17 1BJ, UK}
\altaffiltext{7}{Computational Cosmology Center, Lawrence Berkeley National Laboratory, 1 Cyclotron Road, Berkeley, CA 94720, USA}
\altaffiltext{8}{Department of Astronomy, University of California, Berkeley, CA 94720-3411, USA}
\altaffiltext{9}{Observatories of the Carnegie Institution for Science, 813 Santa Barbara St, Pasadena CA 91101 USA}
\altaffiltext{10}{Department of Astronomy, The Oskar Klein Centre, Stockholm University, AlbaNova University Centre, SE-106 91 Stockholm, Sweden}
\altaffiltext{11}{Dark Cosmology Centre, Niels Bohr Institute, University of Copenhagen, Juliane Maries Vej 30, DK-2100 Copenhagen Ø, Denmark}
\altaffiltext{12}{Spitzer Science Center, MS 314-6, California Institute of Technology, Pasadena, CA 91125, USA}

\begin{abstract}

%        1         2         3         4         5         6         7         8   
%23456789 123456789 123456789 123456789 123456789 123456789 123456789 1234567890

The progenitors of some supernovae (SNe) exhibit 
outbursts with super-Eddington luminosities prior to their final explosions.
This behavior is common among Type IIn SNe, but the driving mechanisms
of these precursors are not yet well understood.
SNHunt\,275 was announced as a possible new SN during May 2015.
Here we report on pre-explosion observations of the location of this event by the
Palomar Transient Factory (PTF)
and report the detection of a precursor about 500\,days prior to
the 2015 May activity (PTF\,13efv).
The observed velocities in the 2015 transient and its 2013 precursor
absorption spectra are low (1000--2000\,km\,s$^{-1}$), so it is not clear yet
if the recent activity indeed marks the final disruption
of the progenitor.
Regardless of the nature of this event, we use the PTF
photometric and spectral observations, as well as
{\it Swift}-UVOT observations, to constrain the efficiency
of the radiated energy relative to the total
kinetic energy of the precursor.
We find that, using an order-of-magnitude estimate and
under the assumption of spherical symmetry,
the ratio of the radiated energy to the kinetic energy
is in the range of $4\times10^{-2}$ to $3.4\times10^{3}$.

\end{abstract}


\section{Introduction}
\label{sec:Introduction}

Some supernova (SN) progenitors exhibit
vigorous variability or possible explosive outbursts
shortly (weeks to years) prior to the SN explosion
(Foley et al. 2007; Pastorello et al. 2007, 2013;
Mauerhan et al. 2013a;
Ofek et al. 2013a, 2013b, 2014a;
Fraser et al. 2013;
Margutti et al. 2014).
Supernovae (SNe) showing such activity
are mostly of Type IIn (and Ibn)
with spectra showing a blue continuum and
hydrogen Balmer (and helium) emission lines
(Schlegel 1990; Filippenko 1991, 1997;
Pastorello et al. 2008; Kiewe et al. 2012).
Moreover, it is possible that other classes of SNe also have precursors
as well (e.g., Corsi et al. 2014; Strotjohann et al. 2015).
Some of the SNe~IIn are presumably powered by conversion of
the large reservoir of kinetic energy to radiated energy
via interaction of the ejecta with circumstellar material
(CSM; e.g., Chugai \& Danziger 1994; Svirski et al. 2012;
Ofek et al. 2014b).
We note that the classification of SNe~IIn is not well-defined;
some SNe display similar spectral features
on timescales of a few days after the explosion, which subsequently
disappear (Niemela et al. 1985;
Fassia et al. 2001;
Gal-Yam et al. 2014;
Shivvers et al. 2015;
Smith et al. 2015;
Khazov et al. 2015;
Yaron et al. 2015).
It is possible that these flash-ionized SNe have lower CSM mass,
and/or a CSM that is confined to short distances from the progenitor
(e.g., Gal-Yam et al. 2014; Groh 2014; Yaron et al. 2015).

Ofek et al. (2014a) systematically searched for pre-explosion
outbursts (precursors) among a sample of 16 SNe~IIn in which the 
hydrogen Balmer lines persist at least until the SN maximum light.
Five possible precursors were found.
Based on this analysis, they conclude that
precursor events among SNe~IIn are common: assuming a 
homogeneous population, at the one-sided 99\% confidence level, more than
$98\%$ of all SNe~IIn have at at least one pre-explosion outburst that is
brighter than $3\times10^{7}$\,L$_\odot$ (absolute magnitude $-14$)
that takes place up to 2.5\,yr prior to the SN explosion. The average rate of such
precursor events during the year prior to the SN explosion is likely
larger than one per year
%$7.5_{-3.6}^{+5.9}$\,yr$^{-1}$
(i.e., multiple events per SN per year), and fainter precursors
are possibly even more common.
% (1$\sigma$ errors).
They also find possible correlations between the integrated luminosity
of the precursor, and the SN total radiated energy, peak luminosity,
and rise time. These correlations are expected if the precursors are
mass ejection events, and the early-time light curve of these SNe is powered
by interaction of the SN ejecta with optically thick CSM.
No precursors were found in a similar search among five SNe~IIn,
recently reported by Bilinski et al. (2015). They do not
provide the absolute-magnitude-dependent search time of their
sample, so direct comparison of the two surveys is not straightforward.

The nature of the SN precursors is unknown,
although several 
theoretical mechanisms have been suggested to explain this
high mass loss in the final stages of stellar
evolution.
These include the pulsational pair instability
(e.g., Rakavy et al. 1967; Woosley et al. 2007; Waldman 2008;
Moriya \& Langer 2015),
bursty shell oxygen burning
(Arnett \& Meakin 2011),
binary evolution (e.g., Chevalier 2012; Soker \& Kashi 2013),
and 
internal gravity waves excited by core convection
(Quataert \& Shiode 2012; Shiode \& Quataert 2013).
In addition to the nature of the engine driving the precursors, another relevant
question is how the mass loss arises and the origin of the radiated luminosity.
In the context of luminous blue variables 
(LBVs) and $\eta$~Carinae in particular,
one can envision mass loss to arise from explosions ---
i.e., shock waves accelerating material at the surface,
later converting the kinetic energy to radiation through the interaction
of the freshly ejected material with previously ejected mass (e.g., Smith 2013).
In this case, we expect the radiated energy
to be less or comparable to the kinetic energy of the ejecta.
In an opposite scenario, a super-Eddington
radiative field drives mass through radiation pressure.
Here we expect the radiated energy to be larger than
the kinetic energy of the ejecta (Shaviv 2000; 2001).

SNHunt\,275 (PSN\,J09093496$+$3307204) was discovered
by Howerton\footnote{Submitted to the CBET confirmation page.}.
Classification of the transient (Elias-Rosa et al. 2015) by the 
Asiago Transient Classification Program using a spectrum taken 
on 2015 Feb 9.93 (UTC dates are used throughout this paper)
revealed a narrow P-Cygni H$\alpha$ line with an emission width of about
900\,km\,s$^{-1}$ and an expansion velocity, derived from the absorption 
component, of 950\,km\,s$^{-1}$.
The P-Cygni profile is superposed on broad H$\alpha$ emission
having a full width at half-maximum intensity (FWHM) of $\sim 6800$\,km\,s$^{-1}$.
Elias-Rosa et al. (2015) also reported on the detection of a possible
source at the transient location in
{\it Hubble Space Telescope} ({\it HST}) images with
apparent magnitudes of 22.8, 21.5, and 22.5 (F606W filter)
on 2009 Feb. 9, 2008 Mar. 30, and 2009 Feb. 25, respectively.
These corresponds to absolute magnitudes of about $-9.7$, $-11.0$,
and $-10.0$, respectively.
Observations on 2015 Mar. 9, Apr. 9, and Apr. 14
showed that the transient brightness had increased
(de Ugarte Postigo et al. 2015a,b,c).
Furthermore, spectroscopic observations on 2015 Apr. 14
(with resolution $R \approx 500$)  did~not detect the P-Cygni
absorption component.
Vinko et al. (2015) reported that
the absolute magnitude of the transient reached $-17$
on 2015 May 18, and suggested that the transient has exploded as a SN.

Here we present PTF observations of the field of this transient
in the years prior to its recent discovery and the detection
of a precursor event reaching an absolute magnitude
of about $-12$ (Duggan et al. 2015).
We use these observations to put limits on the ejected mass
and the radiative efficiency of the precursor.
The radiative efficiency is defined here as the ratio of
the radiated energy to the kinetic energy.
Although the results have an uncertainty of several orders of magnitude,
they provide the foundations for better future measurements.

We assume a distance to the transient of about 30\,Mpc
and a Galactic reddening of $E_{B-V}=0.023$\,mag
(Schlegel et al. 1998).
In \S\ref{sec:Obs} we present our photometric
and spectroscopic observations as well as {\it Swift} observations.
The results are discussed in \S\ref{sec:Disc}
and summarized in \S\ref{sec:Sum}.

\section{Observations}
\label{sec:Obs}

\begin{figure}
\centerline{\includegraphics[width=8.5cm]{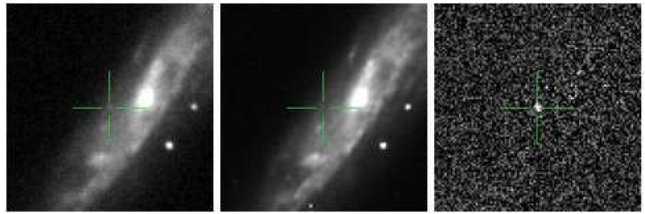}}
\caption{The image-subtraction based detection of PTF\,13efv from
the PTF marshal. From left to right we show the new image,
reference image and subtracted image.
\label{fig:PTF13efv_detection}}
\end{figure}

The Palomar Transient Factory (PTF and iPTF; Law et al. 2009; Rau et al. 2009),
using the 48-inch Oschin Schmidt telescope,
observed the field of SNHunt\,275 starting in March 2009.
On 2013 Dec. 12, PTF detected a new source at
the location of the event, and the transient was named PTF\,13efv
(Figure~\ref{fig:PTF13efv_detection}).
Spectroscopic classification obtained on 2013 Dec. 13
suggested that this is a ``SN imposter'' (e.g., Van Dyk \& Matheson 2012).
All of the PTF observations are reduced using the PTF-IPAC pipeline
(Laher et al. 2014) and the photometric calibration
and magnitude system are described by Ofek et al. (2012a,b).

\begin{figure}
\centerline{\includegraphics[width=8.5cm]{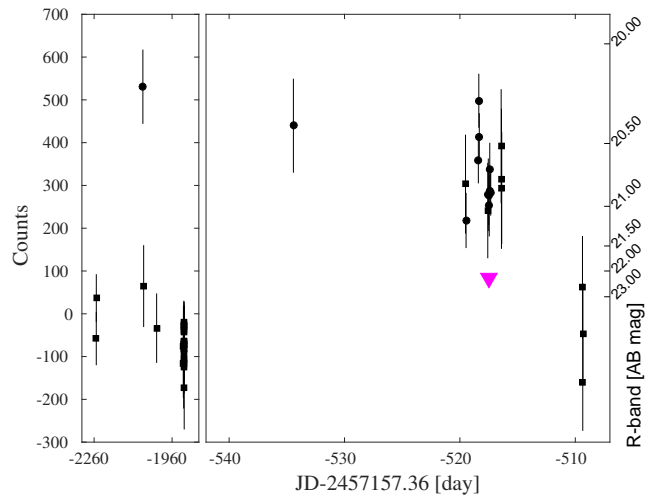}}
\caption{The PTF light curve at the position of SNHunt\,275
prior to its May 2015 event. 
Black filled symbols represent the PTF measurements.
Circles mark individual measurements which are 3 times
above the noise level, while squares represent measurements
which are consistent with 3 times the noise.
The triangle marks a {\it Swift}-UVOT
$UM2$ upper limit.
We note that the weighted mean of the counts during the
2013 outburst is $326\pm18$.
\label{fig:PTF13efv_LC}}
\end{figure}

Photometry of the source was derived using point-spread function
(PSF) fitting photometry
on the subtracted images (see, e.g., Firth et al. 2015 for details).
Three images obtained between 2014 Jan. 23 
and 2014 Apr. 25 were used as a reference.
The PTF $R$-band photometry is listed in
Table~\ref{tab:PTFphot} and the light curve
is presented in Fig~\ref{fig:PTF13efv_LC}.
The $R$-band light curve clearly shows a precursor detected toward the end of
Nov. 2013.
The first detection of this outburst was on 2013 Nov 26.
The next observations, about 2 weeks later, do~not show an indication
for flux variations. Therefore, it is possible that the outburst started
much earlier than Nov 26.
Observations obtained on 2013 Dec 21 indicate
that the source returned to the levels of the reference image.
We note that our PTF $g$-band light curve includes a single
nondetection on 2013 Apr. 22 with a limiting magnitude of 21.1.
The precursor disappeared in the third week of Dec. 2013.
We note that in Figure~\ref{fig:PTF13efv_LC} there is
a single point, on 2009 Sep 10, that looks like an outburst.
In order to test its reality, we ran the newly developed image subtraction
code (Zackay, Ofek, \& Gal-Yam 2016) on the images,
where we constructed a reference image using the optimal image coaddition
algorithm described in Zackay \& Ofek (2015a,b).
This image subtraction code is optimal, in the background dominated
noise limit, and unlike the popular image subtraction methods
it is numerically stable, returns a subtraction image with
uncorrelated noise,
and preserves the shape of cosmic rays and bad pixels.
We found out that the residual causing the detection on
2009 Sep 10 has a sharp shape, indicating that it is likely
a bad pixel or radiation hit event.
Therefore, we conclude it is not an outburst.

We note that there are 21 observations obtained between 2010 Feb 13 and 16.
All these observations have negative fluxes and their weighted mean count is
$-68\pm9$, where the error was estimated using the Bootstrap method
(Efron 1982).
This is likely due to real variability of the progenitor, specifically a decline
in luminosity relative to the reference image.
We note that the formal error on the mean (12 counts) is consistent with the
bootstrap error. This consistency indicates that our error estimate is
reasonable.
For additional tests regarding systematics in
our image subtraction and photometry we refer the reader
to Ofek et al. (2014a).

\begin{deluxetable}{lrcrl}
\tablecolumns{5}
\tablewidth{0pt}
\tablecaption{PTF Photometric Observations}
\tablehead{
\colhead{MJD}           &
\colhead{Counts}        &
\colhead{Counts error}  &   
\colhead{Mag}           &
\colhead{Mag err.}  \\
\colhead{(day)}       &
\colhead{}            &
\colhead{}            &
\colhead{(mag)}       &
\colhead{(mag)}
}
\startdata
54905.1661 &$      -57.7$ &     62.5 & $ >21.32$ & \nodata  \\
54905.2811 &$       36.5$ &     56.0 & $ >21.44$ & \nodata  \\
55084.5089 &$      530.7$ &     86.8 & $ 20.19 $ & 0.18     \\
55087.5125 &$       64.7$ &     95.7 & $ >20.85$ & \nodata  \\
55137.4415 &$      -33.6$ &     81.1 & $ >21.03$ & \nodata  \\
55240.1202 &$     -116.8$ &     79.8 & $ >21.05$ & \nodata  \\
55240.5060 &$      -76.5$ &     93.7 & $ >20.88$ & \nodata  \\
55241.1410 &$     -116.6$ &     60.3 & $ >21.36$ & \nodata  \\
55241.4996 &$     -125.0$ &     96.4 & $ >20.85$ & \nodata  \\
55242.1389 &$      -20.3$ &     50.7 & $ >21.54$ & \nodata  \\
55242.4967 &$      -42.6$ &     68.8 & $ >21.21$ & \nodata  \\
55243.1236 &$      -72.9$ &     58.5 & $ >21.39$ & \nodata  \\
55243.1739 &$      -70.5$ &     49.3 & $ >21.58$ & \nodata  \\
55243.1756 &$      -25.3$ &     54.2 & $ >21.47$ & \nodata  \\
55243.2207 &$      -66.8$ &     49.3 & $ >21.58$ & \nodata  \\
55243.2223 &$     -109.3$ &     51.1 & $ >21.54$ & \nodata  \\
55243.2676 &$      -29.3$ &     49.5 & $ >21.57$ & \nodata  \\
55243.2694 &$      -82.7$ &     49.1 & $ >21.58$ & \nodata  \\
55243.3150 &$      -72.8$ &     50.9 & $ >21.54$ & \nodata  \\
55243.3167 &$      -28.5$ &     47.1 & $ >21.62$ & \nodata  \\
55243.3610 &$      -38.1$ &     51.8 & $ >21.52$ & \nodata  \\
55243.3628 &$      -81.8$ &     50.5 & $ >21.55$ & \nodata  \\
55243.4087 &$      -94.1$ &     54.3 & $ >21.47$ & \nodata  \\
55243.4105 &$      -62.8$ &     53.9 & $ >21.48$ & \nodata  \\
55243.4544 &$     -105.5$ &     80.8 & $ >21.04$ & \nodata  \\
55243.5096 &$     -173.2$ &     97.2 & $ >20.84$ & \nodata  \\
56622.4310 &$      439.6$ &    109.9 & $ 20.39 $ & 0.27 \\
56637.3456 &$      303.0$ &    115.6 & $ >20.65$ & \nodata  \\
56637.3942 &$      218.1$ &     64.5 & $ 21.15 $ & 0.32 \\
56638.4580 &$      359.6$ &     54.8 & $ 20.61 $ & 0.17 \\
56638.4985 &$      498.0$ &     63.1 & $ 20.26 $ & 0.14 \\
56638.5411 &$      412.3$ &     56.6 & $ 20.46 $ & 0.15 \\
56639.2757 &$      241.3$ &    111.5 & $ >20.69$ & \nodata  \\
56639.3197 &$      278.1$ &     84.9 & $ 20.89 $ & 0.33     \\
56639.3600 &$      277.3$ &     69.6 & $ 20.89 $ & 0.27     \\
56639.4031 &$      252.7$ &     71.8 & $ 20.99 $ & 0.31     \\
56639.4473 &$      336.7$ &     63.0 & $ 20.68 $ & 0.20     \\
56639.4895 &$      287.7$ &     52.9 & $ 20.85 $ & 0.20     \\
56639.5333 &$      283.7$ &     54.3 & $ 20.87 $ & 0.21     \\
56640.4351 &$      391.6$ &    133.3 & $ >20.50$ & \nodata  \\
56640.4606 &$      315.5$ &    163.7 & $ >20.27$ & \nodata  \\
56640.4854 &$      293.8$ &    130.7 & $ >20.52$ & \nodata  \\
56647.4823 &$       61.8$ &    120.0 & $ >20.61$ & \nodata  \\
56647.5074 &$     -159.5$ &    114.1 & $ >20.66$ & \nodata  \\
56647.5221 &$      -46.0$ &    117.0 & $ >20.64$ & \nodata  
\enddata
\tablecomments{Image-subtraction-based $R$-band
photometry of PTF\,13efv. MJD is the modified Julian day.
AB Magnitudes are presented as lower limits when the detection is
less than 3$\sigma$ than the noise level.
The count rate can be converted to AB magnitude
with $M=27-2.5\log_{10}({\rm Counts})$.}
\label{tab:PTFphot}
\end{deluxetable}

\begin{figure}
\centerline{\includegraphics[width=8.3cm]{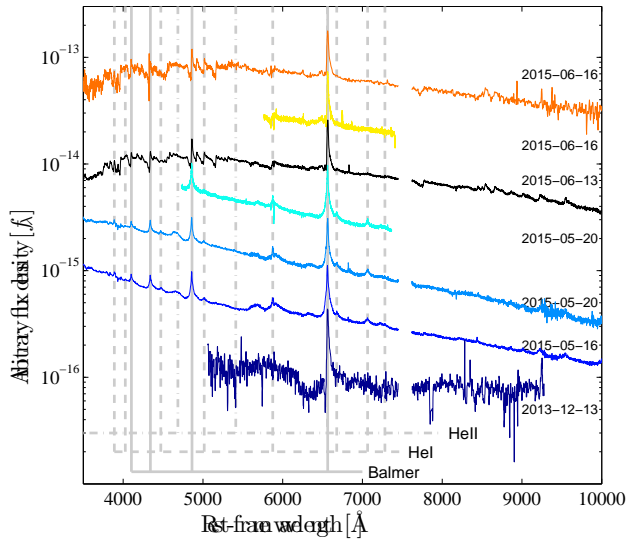}}
\caption{Spectra of PTF\,13efv/SNHunt\,275 obtained
as part of the PTF project. The spectra are corrected for the
host-galaxy redshift.
The 2013 Dec. 13 spectrum is smoothed using a five-pixel median filter.
Telluric line regions were removed from the spectra.
\vspace{0.2cm}
\label{fig:SpecAll_PTF13efv}}
\end{figure}

\begin{figure}
\centerline{\includegraphics[width=8.3cm]{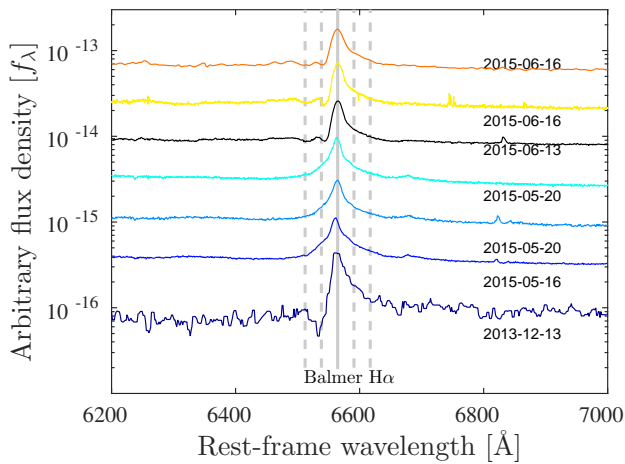}}
\caption{Close-up view of the H$\alpha$ region of the PTF spectra (Fig.~\ref{fig:SpecAll_PTF13efv}).
The solid vertical line represents the rest-frame wavelength of the
H$\alpha$ line, while the dashed lines are for velocities
of 1000 and 2000\,km\,s$^{-1}$.
On June 2015 a double absorption P-Cygni profile, with velocities of 1000
and 2000\,km\,s$^{-1}$, appears.
\label{fig:SpecHa_PTF13efv}}
\end{figure}

  Most of the optical spectra (see Table~\ref{tab:Spec}) 
were obtained with the Low Resolution
Imaging Spectrometer (LRIS; Oke et al. 1995) on the Keck-1 10\,m
telescope, although a few spectra were also taken with the DEep
Imaging Multi-Object Spectrograph (DEIMOS; Faber et al. 2003) on the
Keck-2 10\,m telescope, the Kast spectrograph (Miller \& Stone 1993) 
on the Shane 3\,m telescope at Lick Observatory, and
the Gemini-North Multiobject Spectrograph (GMOS; Hook et al. 2004) 
on the 8\,m Gemini-N telescope. Spectral
reductions followed standard techniques (e.g., Matheson et al. 2000;
Silverman et al. 2012). All spectra are publicly available online via the
Weizmann Interactive Supernova Data Repository, 
WISeREP\footnote{http://wiserep.weizmann.ac.il/}
(Yaron \& Gal-Yam 2012).
The spectra are presented in Figure~\ref{fig:SpecAll_PTF13efv} and a 
close-up view of the H$\alpha$ line
is shown in Figure~\ref{fig:SpecHa_PTF13efv}.

The first spectrum was obtained during the
Dec. 2013 outburst.
It exhibits a strong and narrow H$\alpha$ emission line
(FWHM $\ltorder 500$\,km\,s$^{-1}$)
with a narrow P-Cygni absorption component at a velocity of $\sim1300$\,km\,s$^{-1}$
(measured relative to the peak of the emission line).
The spectrum continuum is consistent with an effective temperature
of about 5750\,K and a radius of $\sim 4\times10^{14}$\,cm (see Table~\ref{tab:Spec}).
We note that blueward of the H$\alpha$ line, there is a minor decrement
in the flux level. If this is due to a P-Cygni profile,
in addition to the narrow P-Cygni at 1300\,km\,s$^{-1}$,
then this indicates velocities of up to 15,000\,km\,s$^{-1}$.
However, the nature of this decrement is not clear.
The H$\alpha$ luminosity at this epoch is
roughly $1.2\times10^{39}$\,erg\,s$^{-1}$.

After the May 2015 rebrightening, the spectra become bluer, and the
H$\alpha$ emission line in the Keck/DEIMOS spectrum
is well described by a two-component Gaussian
with component widths of $\ltorder500$\,km\,s$^{-1}$ and $\sim2000$\,km\,s$^{-1}$.
A month later, the spectra become redder 
and two P-Cygni absorption features
are detected in all of the Balmer lines:
one with a velocity of $\sim1000$\,km\,s$^{-1}$
(as before),
and a new absorption feature with a velocity of $\sim2000$\,km\,s$^{-1}$.
We note that the DEIMOS spectrum shows the Na\,I absorption doublet
(5890, 5896\,\AA) 
at zero redshift and at the host-galaxy redshift.
The equivalent width of the Na\,I doublet at the host-galaxy redshift
is about 2.3 times stronger than the
Galactic Na~I absorption line.
Therefore, it is likely that there is host-galaxy extinction
in the direction of this event.

The H$\alpha$ line luminosity as measured in the Keck/DEIMOS
spectrum on 2015 May 20 is about $1.2\times10^{40}$\,erg\,s$^{-1}$.
This is over an order of magnitude stronger than the luminosity during the 2013 outburst.
We verified the flux calibration
is correct by calculating the $V_{{\rm UVOT}}$-band synthetic magnitude
from the spectrum and comparing it with the {\it Swift}-UVOT
photometry.

\begin{deluxetable*}{llllll}
\tablecolumns{6}
\tablewidth{0pt}
\tablecaption{Log of Spectroscopic Observations}
\tablehead{
\colhead{Telescope}     &
\colhead{Instrument}     &
\colhead{Setup}          &
\colhead{MJD}            &
\colhead{Temp.}          &
\colhead{Radius}         \\
\colhead{}               &
\colhead{}               &
\colhead{}               &
\colhead{}               &
\colhead{(K)}              &
\colhead{(cm)}     
}
\startdata
Gemini-N  & GMOS        & R400/G5305        & 56639.8 & 5820  & $4\times10^{14}$ \\
Keck-I    & LRIS        & 400/3400, 400/8500 & 57158.3 & 10,800 & $6\times10^{14}$ \\
Keck-I    & LRIS        & 400/3400, 400/850  & 57162.3 & 9230  & $4\times10^{14}$ \\
Keck-II   & DEIMOS      & 1200G             & 57162 & 9030  & $7\times10^{14}$ \\
Keck-I    & LRIS        & 400/3400, 400/8500 & 57186.3 & 6010  & $1\times10^{15}$ \\ 
Keck-I    & LRIS        & 600/4000, 1200/7500& 57189 & \nodata  & \nodata \\ 
Lick 3\,m  & Kast        & 600/4310, 300/7500 & 57191 & 5960  & $1\times10^{15}$ 
\enddata
\tablecomments{MJD is the modified Julian day.
The temperature and radius are based on fitting a black-body
continuum to the spectra (excluding the H$\beta$ and H$\alpha$ regions).
Since the temperatures may be affected by
metal absorption, they should be regarded as lower limits.
Similarly, the radii should be regarded as upper limits.
``Setup'' indicates the grating name, or (respectively) the blue grism 
and red grating.
The spectra were obtained at the parallactic angle, and were corrected
for airmass-effects using the mean atmospheric extinction curve for
each site.
We note that the Galactic reddening ($E_{B-V}=0.023$\,mag)
is taken into account in the effective temperature calculations.
However, we ignored the unknown host extinction.
If the host extinction is indeed a factor of 2 larger than the Galactic
extinction, as suggested by the Na\,I absorption doublet,
then the lower limit on the effective temperature will be 300 to 1000\,K
higher than listed in the Table.
\label{tab:Spec}}
\end{deluxetable*}

\begin{figure}
\centerline{\includegraphics[width=8.5cm]{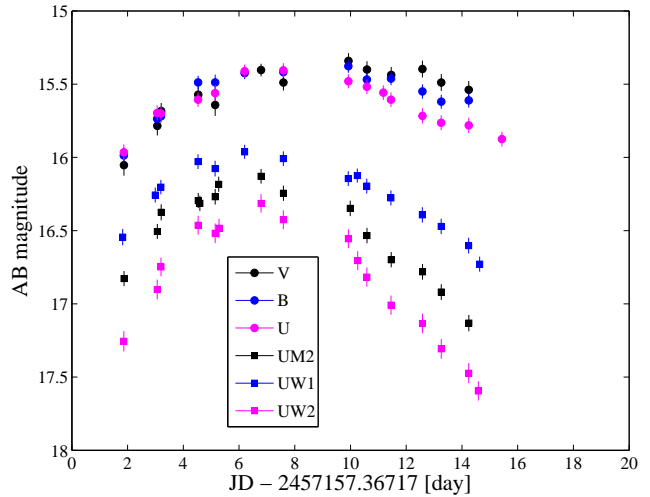}}
\caption{The {\it Swift}-UVOT apparent magnitude
light curves (not corrected for extinction)
in the $UW2$, $UW1$, $UM2$, $U$, $B$, and $V$ bands.
The host contributions, estimated based on images taken prior to
Jan. 2015, are subtracted (see Table~\ref{tab:UVOTphot} caption).
Precursors are not detected in observations prior to the May 2015 event.
The full list of photometric measurements from 2008 until 2015 appears in
Table~\ref{tab:UVOTphot}.
\label{fig:SwiftUVOT_LC}}
\end{figure}

SNHunt\,275 exploded in NGC\,2770, which has been the home
of several SNe (e.g., Th{\"o}ne et al. 2009),
among which was SN\,2008D (Soderberg et al. 2008; Modjaz et al 2009).
Thus, the host galaxy has been observed many times
and by various instruments.
Specifically, since 2008, it was observed
by the Ultra-Violet/Optical Telescope (UVOT; Roming et al. 2005)
onboard the {\it Swift} satellite (Gehrels et al. 2004).
%The 2015 event observations were triggered by
%Brown,
%Campana,
%Postigo,
%Kuin, and Margutti.
Some of these observations have already been reported
(e.g., Campana et al. 2015).
The data were reduced using standard procedures (e.g., Brown et al. 2009).
Flux from the transient was extracted from a $3''$-radius aperture, with a correction applied
to transform the photometry on the standard UVOT system (Poole et al. 2008).
The resulting measurements, all of which have been converted to the AB system (Oke \& Gunn 1983), are listed in
Table~\ref{tab:UVOTphot} and displayed in Figure~\ref{fig:SwiftUVOT_LC}.
Since there are no UVOT detections of the object prior to
$t_{0}$ ($=2457157.36$, see definition below),
Figure~\ref{fig:SwiftUVOT_LC} shows only
measurements taken after $t_{0}$.
We note that the contribution from the host galaxy was subtracted by
removing the 
(coincidence-loss corrected)
mean count rate observed prior to January 2015.

\begin{deluxetable}{llrl}
\tablecolumns{4}
\tablewidth{0pt}
\tablecaption{{\it Swift}-UVOT Photometric Observations}
\tablehead{
\colhead{Filter}       &
\colhead{JD\,$-\,t_{0}$}    &
\colhead{Counts}        &
\colhead{Counts error}  \\
\colhead{}              &
\colhead{(day)}           &
\colhead{}            &
\colhead{}          
}
\startdata
     $V$  &$  -2685.8316 $&   $-0.016$ &    0.054 \\ %& 27.64 &  0.07\\
     $V$  &$  -2682.2874 $&    0.087 &    0.074 \\ %& 20.53 &  0.08\\
     $V$  &$  -2680.7415 $&   $-0.007$ &    0.040 \\ %& 26.77 &  0.05\\
     $V$  &$  -2679.6697 $&    0.004 &    0.052 \\ %& 23.76 &  0.06\\
     $V$  &$  -2678.8016 $&    0.000 &    0.047    %& 25.00 &  0.06
\enddata
\tablecomments{Time is given relative to 
$t_{0}=$ 2,457,157.36.
The counts are background subtracted,
where the background is estimated as the mean of all the observations
in a given filter obtained before 2015 Jan. 1.
The subtracted backgrounds are
0.892, 1.496, 0.743, 0.085, 0.206, and 0.146 counts
for the $V$, $B$, $U$, $UVM2$, $UVW1$, and $UVW2$ filters, respectively.
The zero points to convert these counts to AB magnitudes
are 17.88, 18.99, 19.36, 18.97, 18.53, and 19.07 mag,
for the $V$, $B$, $U$, $UVW1$, $UVM2$, and $UVW2$ filters,
respectively.
This table is published in its entirety in the electronic edition of
{\it ApJ}. A portion of the full table is shown here for
guidance regarding its form and content.
\label{tab:UVOTphot}}
\end{deluxetable}

\begin{figure}
\centerline{\includegraphics[width=8.5cm]{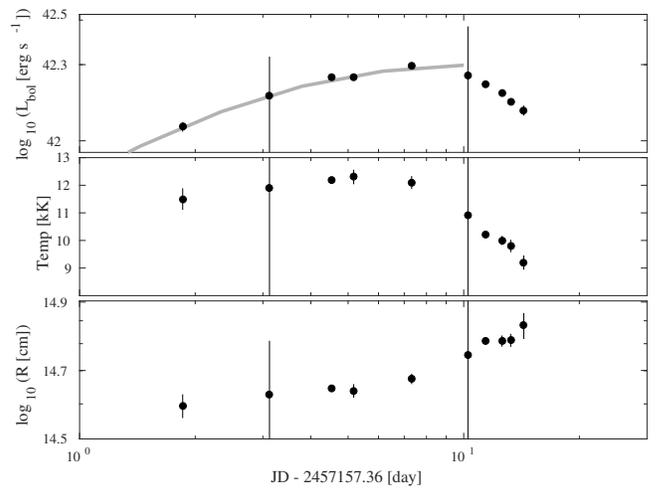}}
\caption{Fitted bolometric luminosity (upper panel),
effective temperature (middle panel),
and effective radius (lower panel) as a function of time
since the fitted $t_{0}$. See text for details.
The gray line shows the best fit exponential rise timescale.
\label{fig:UVOT_LRT}}
\end{figure}

\begin{figure}
\centerline{\includegraphics[width=8.5cm]{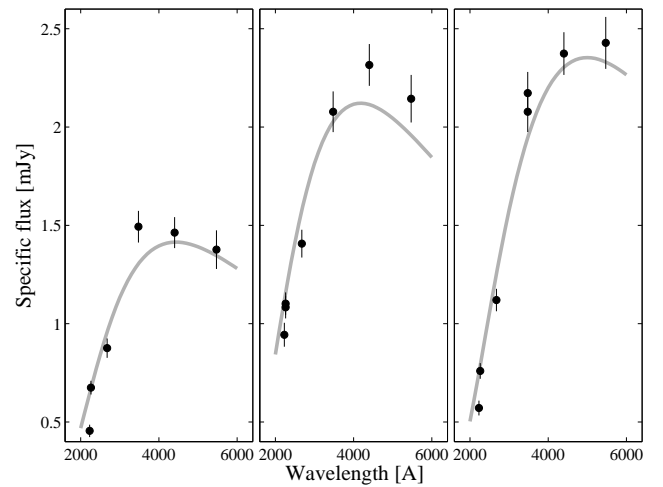}}
\caption{From left to right, we present the UVOT spectral energy distribution
of SN\,Hunt 275, on three epochs: 1.8, 4.5 and 14.3\,days after $t_{0}$, respectively.
The gray lines represent the best-fit black-body curve.
\label{fig:UVOT_SED}}
\end{figure}

We used the UVOT observations to construct the bolometric
light curve of the transient. This was done by correcting the
measurements for Galactic reddening of $E_{B-V}=0.023$\,mag
(Schlegel et al. 1998; Cardelli et al. 1989), and fitting a black-body continuum to all of
the observations in one-day bins (only in bins having observations in more than three bands).
The fitted bolometric light curve, effective temperature,
and radius are presented in Figure~\ref{fig:UVOT_LRT},
while the fitted measurements are listed in Table~\ref{tab:UVOTbol}.
Figure~\ref{fig:UVOT_SED} presents the UVOT spectral energy distribution,
along with the best-fit black-body curve,
on three epochs, 1.8, 4.5 and 14.3\,days after $t_{0}$.
The uncertainties were estimated using the bootstrap method (Efron 1982)
applied to each time bin.
Following Ofek et al. (2014c), we further estimated the rise timescale of the event
by fitting the luminosity ($L$) with an exponential rise of the form
\begin{equation}
L=L_{{\rm max}}(1-\exp[-(t-t_{0})/t_{{\rm rise}}]),
\label{eq:rise}
\end{equation}
where $L_{{\rm max}}$ is the fitted maximum luminosity
and $t_{0}$ is the fitted time of zero flux.
We fitted the first four detections and
estimate that $t_{{\rm rise}}\approx2.2\pm1.6$\,days,
$L_{{\rm max}}=(2.0\pm0.3)\times10^{42}$\,erg\,s$^{-1}$,
and $t_{0}=$ 2,457,157.36 $\pm 2.2$\,day.
We note that $t_{0}$ does~not necessarily mark the time of explosion.
\begin{deluxetable}{llcl}
\tablecolumns{4}
\tablewidth{0pt}
\tablecaption{{\it Swift}-UVOT bolometric light curve}
\tablehead{
\colhead{JD\,$-\,t_{0}$}    &
\colhead{$L_{{\rm bol}}$} &
\colhead{Temp.}        &
\colhead{Radius}       \\
\colhead{(day)}           &
\colhead{($10^{42}$\,erg\,s$^{-1}$)}  &
\colhead{(K)}             &
\colhead{($10^{14}$\,cm)}         
}
\startdata
   1.860 &$  1.14\pm0.05$ &$  11,500\pm600$&$   4.0\pm0.5$\\
   3.126 &$  1.50\pm0.03$ &$  11,900\pm300$&$   4.3\pm0.2$\\
   4.540 &$  1.78\pm0.02$ &$  12,200\pm100$&$   4.4\pm0.1$\\
   5.176 &$  1.79\pm0.36$ &$  12,300\pm600$&$   4.4\pm1.7$\\
   7.325 &$  1.98\pm0.04$ &$  12,100\pm200$&$   4.7\pm0.2$\\
  10.260 &$  1.81\pm0.03$ &$  10,900\pm200$&$   5.5\pm0.2$\\
  11.416 &$  1.67\pm0.04$ &$  10,200\pm200$&$   6.1\pm0.3$\\
  12.582 &$  1.54\pm0.04$ &$  10,000\pm300$&$   6.1\pm0.4$\\
  13.257 &$  1.43\pm0.04$ &$   9800\pm200$&$   6.1\pm0.3$\\
  14.335 &$  1.32\pm0.05$ &$   9200\pm300$&$   6.8\pm0.5$
\enddata
\tablecomments{Bolometric luminosity, effective temperature, and radius
estimated from a black-body fit to the {\it Swift}-UVOT observations
(Table~\ref{tab:UVOTphot}) corrected for Galactic extinction.
Like in Table~\ref{tab:Spec}, the temperature measurements should be regarded
as lower limits on the effective temperature.
Assuming the host extinction is twice as large as the Galactic extinction
(i.e., as suggested by the Na\,I absorption doublet)
the lower limit on the temperature will be higher by up to about 1000\,K.
\label{tab:UVOTbol}}
\end{deluxetable}
%

%--- Paragraph on Swift-XRT observations --
For each {\em Swift}-XRT image of the transient,
we extracted the number of X-ray counts in the 0.2--10\,keV
band within an aperture of $9''$ radius
centered on the transient position.
This aperture contains $\sim 50$\% of
the source flux (Moretti et al.\ 2004).
The background count rates were estimated in
an annulus around the transient
location, with an inner (outer) radius of $50''$ ($100''$).
The log of {\it Swift}-XRT observations,
along with the source and background X-ray counts in the individual
observations, is given in Table~\ref{tab:XRTobs}.
While binning the observations in ten-day bins, we did~not
detect X-rays from this position with a false-alarm probability lower than 4\%.
In the two weeks after $t_{0}$, we can set a 2$\sigma$ upper limit
of $0.26$\,count\,ks$^{-1}$ in the 0.2--10\,keV range.
Assuming a Galactic neutral hydrogen
column density of $n_{{\rm H}}=1.8\times10^{20}$\,cm$^{-2}$
and an intrinsic power-law spectrum with a photon index of $\Gamma=2$,
this translates to an upper limit on the luminosity
of $L_{X}<1.1\times10^{39}$\,erg\,s$^{-1}$ within the {\it Swift}-XRT energy range.
\begin{deluxetable}{lrrr}
\tablecolumns{4}
\tablewidth{0pt}
\tablecaption{{\it Swift}-XRT observations}
\tablehead{
\colhead{JD\,$-\,t_{0}$}          &
\colhead{Exp. time}          &
\colhead{Source}      &
\colhead{Background}     \\
\colhead{(day)} &
\colhead{(s)} &
\colhead{(counts)}            &
\colhead{(counts)}
}
\startdata
$-2685.835$ &  9595.4 &   2 &  38\\
$-2682.295$ &  4563.3 &   1 &  55\\
$-2680.750$ & 28649.6 &   3 & 124\\
$-2679.740$ & 11428.2 &   2 &  28\\
$-2678.811$ & 15785.9 &   0 &  65
\enddata
\tablecomments{Source is the number of counts in 
a $9''$-radius aperture of the source position
and in the 0.2--10\,keV band.
Background is the number of counts,
in the 0.2--10\,keV band,
in an annulus of inner (outer)
radius of $50''$ ($100''$) around the source.
The ratio between the background annulus area and the aperture area
is $92.59$.
This table is published in its entirety in the electronic edition of
{\it ApJ}. A portion of the full table is shown here for
guidance regarding its form and content.}
\label{tab:XRTobs}
\end{deluxetable}

Throughout this paper we assume a distance to SNHunt\,275 of
30\,Mpc (distance modulus 32.38\,mag).
The reduction and analysis presented here is based mainly on
tools available as part of the MATLAB astronomy and astrophysics package (Ofek 2014).

\section{Discussion}
\label{sec:Disc}

Here we briefly review the properties of the 2013 event (\S\ref{sec:re}),
and discuss the question of whether SNHunt\,275 marks the final disruption
of the star (\S\ref{sec:Nat}).
Furthermore, by analyzing
the properties of the 2013 precursor
and the latest explosion
(May 2015), we attempt to constrain the physical setup of this explosion,
and specifically the radiative efficiency of the precursor explosions (\S\ref{sec:Rad}).
In \S\ref{sec:What} we discuss the question of whether the possible mass loss is driven 
by a radiation field, or the radiation is
generated by the mass-loss interaction with previously emitted
material.

\subsection{The 2013 event}
\label{sec:rev}

To summarize -- the 2013 outburst took place about 500\,days
prior to the May 2015 main event and reached a peak
absolute magnitude of about $-11.9$ in $R$-band ($\approx 1.7\times10^{40}$\,erg\,s$^{-1}$).
The duration of this outburst was longer than 20\,days, hence
the integrated radiated energy in $R$-band is $>2.4\times10^{46}$\,erg.
An interesting fact is that the outbursts decayed fast, on less than a week time scale.
A spectrum taken during the outburst revealed Balmer lines with P-Cygni
profile with a velocity of about 1000\,km\,s$^{-1}$.
These properties are summarized in Table~\ref{tab:SNprop}.

In terms of peak absolute magnitude, and the total radiated energy
this event is
at the low end of the precursor event population reported
in Ofek et al. (2014a).
However, this is a clear selection bias.
One of the most well studied SN
showing multiple precursor events is SN\,2009ip
(Smith et al. 2010; Mauerhan et al. 2013;
Ofek et al. 2013b; Pastorello et al. 2013;
Prieto et al. 2013; Fraser et al. 2015).
Interestingly, SN\,2009ip likely showed four
events, prior to its presumably final explosion
on September 2012 (e.g., Smith et al. 2010).
These events took place at about $-25$,
$-660$, $-710$ and $-1060$\,day prior to the
latest explosion.
The activity of SNHunt\,275 on time scales of tens 
to hundreds of days prior to the presumably final explosion
is similar to the one observed in SN\,2009ip.
One difference is that the peak luminosity of the outbursts
seen in SN\,2009ip was about an order of magnitude higher
than that of SNHunt\,275.
One key question that is not yet clear in the cases
of SN\,2009ip and SNHunt\,275 is if we saw the final death of the star,
or the latest events are just other, brighter than average, outbursts.

\subsection{The Nature of the 2015 Event}
\label{sec:Nat}

A close-up view of the evolution of the H$\alpha$ line is presented
in Figure~\ref{fig:SpecHa_PTF13efv}.
An interesting fact is the appearance of a single P-Cygni absorption
feature during the 2013 outburst, and two absorption
features in spectra taken about one month after the May 2015 event.
This double P-Cygni absorption feature,
with $\sim1000$ and $\sim2000$\,km\,s$^{-1}$ velocities,
is seen in all of the Balmer lines in the spectrum.

This can be interpreted in several ways; 
here we mention two obvious possibilities.
First, the absorption at 2000\,km\,s$^{-1}$ can be produced by material ejected
after the 2013 outburst but before the May 2015 event (e.g., during the Feb. 2015 rebrightening).
At early times during the May 2015 event,
the 2000\,km\,s$^{-1}$ gas is hot and generates
the 2000\,km\,s$^{-1}$ wide emission lines.
Later on this relatively dense gas cools,
%(faster than the slower and thinner gas at large distances),
and we detect a P-Cygni profile with two absorption features at
1000 and 2000\,km\,s$^{-1}$.
In this case, the 1000\,km\,s$^{-1}$ absorption
is likely tracing material ejected during the 2013 event, or earlier.
If the star exploded as a SN in May 2015 (with velocities of about 10,000\,km\,s$^{-1}$),
the SN ejecta will reach
the CSM shells after a few weeks.
%In this case, we expect to detect broad lines in the near future, and perhaps X-ray and radio emission.

Alternatively, it is possible that the May 2015 event released
material at 2000\,km\,s$^{-1}$
that at early times is seen in emission and later in absorption.
If this scenario is correct, we predict that X-ray and radio emission will not be detected,
since the shock velocity is too low.
The current X-ray nondetection,
the reddening (cooling) of the spectra, and lack of broad
spectral features suggest that the star has not exploded yet.

The scenarios we discuss do~not cover all possibilities.
For example, breaking the spherical symmetry
gives rise to a large number of scenarios.
However, these are very hard to constrain given the limited information
at hand.
We stress that the observations are hardly conclusive, and it is still
not clear what the nature of the May 2015 transient is.
We note that future {\it HST} observations, taken after the transient light
fades away, may check if the progenitor is still visible, and hence
whether the May 2015 event marked the final explosion of the star.

\subsection{The Radiative Efficiency of Precursors}
\label{sec:Rad}

Table~\ref{tab:SNprop} lists the measured properties of the
precursor and the possible SN explosion.
These properties were estimated based on the PTF light curve,
the spectra, the {\it Swift}-UVOT data,
and the {\it HST} observations (Elias-Rosa et al. 2015).
Next, we use these properties to estimate the radiative efficiency
of the precursor.
\begin{deluxetable*}{ll}
\tablecolumns{2}
\tablewidth{0pt}
\tablecaption{PTF\,13efv and SNHunt\,275 Observed Properties}
\tablehead{
\colhead{Property}          &
\colhead{Value}
}
\startdata
Progenitor luminosity ({\it HST} observations)               & $(2-7)\times10^{39}$\,erg\,s$^{-1}$ \\
\hline
Precursor $R$-band peak luminosity                     & $1.7\times10^{40}$\,erg\,s$^{-1}$ \\
Precursor peak $R$-band absolute mag                & $-11.9$\,mag \\
Precursor total integrated radiated $R$ band           & $>2.4\times10^{46}$\,erg \\
Precursor duration                                     & $>20$\,days \\
Precursor decay timescale                              & $<7$\,days \\
Precursor time before the explosion                    & $\sim 500$\,days \\
Precursor velocity from P-Cygni profile                & $\sim 1000$\,km\,s$^{-1}$ \\
$UM2-R$ (AB) color index                               & $>1.4$\,mag \\
\hline
May 2015 event rise timescale                          & $2.2\pm1.6$\,days\\
May 2015 event peak bolometric luminosity              & $\sim2\times10^{42}$\,erg\,s$^{-1}$\\
May 2015 event integrated bolometric radiation         & $\sim1.8\times10^{48}$\,erg
\enddata
\tablecomments{Properties of the progenitor (upper block),
December 2013 precursor event (middle block),
and May 2015 event (lower block).
The blocks are separated by horizontal lines.
\label{tab:SNprop}}
\end{deluxetable*}
The goal of this section is to roughly estimate the CSM mass
ejected in the 2013 outburst, and
to estimate the ratio between the radiated luminosity
and kinetic energy of the precursor.
This measurement has the potential to resolve the key question:
what drives the CSM ejection?
For example, a ratio much smaller than one,
favors models in which the radiation is generated
by conversion of the kinetic energy of the ejected mass to
radiation via interaction (forming collisonless shocks;
e.g., Katz et al. 2011; Murase et al. 2011; 2014)
with previously emitted material,
over models in which a super-Eddington luminosity drives the ejection
of the CSM.

Since the precursor has super-Eddington luminosity (for a $\ltorder 100$\,M$_{\odot}$ progenitor),
it is likely that the outburst was associated with mass ejection.
Here we attempt to 
estimate the physical parameters
of the precursor (e.g., ejected mass) and to use
it to estimate the ratio of the
radiated energy of the precursor to its kinetic energy,
which we call the {\it precursor radiative efficiency}:
%\note{(Use $\left\langle L \right\rangle$ instead of $<L>$)}
\begin{eqnarray}
\epsilon_{R} \equiv & \frac{2\int{Ldt}}{M_{{\rm CSM}} v_{{\rm CSM}}^{2}} \cr
            \approx   & 3\times10^{-3}
                     \Big(\frac{\left\langle L \right\rangle}{1.7\times10^{40}\,{\rm erg\,s}^{-1}}\Big) 
                     \Big(\frac{\Delta{t}}{20\,{\rm day}}\Big) \cr
            \times & \Big(\frac{M_{{\rm CSM}}}{{\rm M}_{\odot}}\Big)^{-1}
                     \Big(\frac{v_{{\rm CSM}}}{1000\,{\rm km\,s}^{-1}}\Big)^{-2}.
\label{eq:epsR}
\end{eqnarray}
Here $L$ is the precursor luminosity as a function of time $t$,
$\Delta{t}$ is the precursor duration,
$M_{{\rm CSM}}$ is the precursor ejecta mass,
and $v_{{\rm CSM}}$ is the precursor ejecta velocity.
The CSM mass can be expressed as
\begin{equation}
M_{\rm CSM} = \Big(\frac{1}{\epsilon_{R}}\Big) \Big(\frac{2\left\langle L \right\rangle \Delta{t}}{v_{\rm CSM}^{2}}\Big).
\label{eq:M}
\end{equation}

The radiative efficiency allows us to
relate between the observed luminosity integrated over time,
the CSM velocity, and mass.
Furthermore, the exact value of the
radiative efficiency likely depends on the
CSM ejection mechanism, and therefore it may be useful for
testing some theoretical ideas regarding the precursor physical mechanism
(see \S\ref{sec:What}).

However, our derivation is an order-of-magnitude estimate that
relies on several assumptions, which are not necessarily correct.
For example, we assume that the CSM has spherical
symmetry and is not heavily clumped.
Nevertheless, as far as we know, this is the only existing estimate
for the radiative efficiency of a precursor.

The distance the precursor ejecta can travel during its
20-days ejection is
\begin{eqnarray}
r_{{\rm CSM}}\approx & v_{{\rm CSM}}\Delta{t} \cr
          \approx   & 1.7\times10^{14}
                    \Big(\frac{v_{{\rm CSM}}}{1000\,{\rm km\,s}^{-1}}\Big)
                    \Big(\frac{\Delta{t}}{20\,{\rm day}}\Big)\,{\rm cm},
\label{eq:rcsm}
\end{eqnarray}
where $\Delta{t}$ is the duration of the precursor ($\ge20$\,day).
An order-of-magnitude estimate of the mean density of the
ejected CSM (during its ejection) is
\begin{eqnarray}
n\approx & \frac{M_{{\rm CSM}}}{\mu_{{\rm p}} m_{{\rm p}} 4/3 \pi r_{{\rm CSM}}^{3} } \cr
    \approx  & 2.7\times10^{11}
                          \Big(\frac{1}{\epsilon_{R}}\Big)
                          \Big(\frac{\left\langle L \right\rangle}{1.7\times10^{40}\,{\rm erg\,s}^{-1}}\Big) 
                          \Big(\frac{\mu_{{\rm p}}}{0.6}\Big)^{-1} \cr
    \times &              \Big(\frac{\Delta{t}}{20\,{\rm day}}\Big)^{-2}
                          \Big(\frac{v_{{\rm CSM}}}{1000\,{\rm km\,s}^{-1}}\Big)^{-5}\,{\rm cm}^{-3}.
\label{eq:n}
\end{eqnarray}
Here $\mu_{{\rm p}}$ is the mean molecular weight (assumed to be 0.6).

Another constraint comes from the fact that the precursor
radiation disappeared on a timescale shorter than one week (Figure~\ref{fig:PTF13efv_LC});
thus, the cooling timescale is $\lesssim1$ week.
The Bremsstrahlung cooling timescale, which gives an upper limit on the cooling 
timescale, is given by
\begin{equation}
t_{{\rm cool}} \ltorder 1.76\times10^{13} \Big(\frac{T}{10^{4}\,{\rm K}}\Big)^{1/2}
  \Big(\frac{n}{1\,{\rm cm}^{-3}}\Big)^{-1}
  \Big(\frac{Z}{1}\Big)^{-2}\,{\rm s},
\label{eq:tcool}
\end{equation}
where $T$ is the gas temperature and $Z$ is the atomic number (number of protons),
and this can be translated to a lower
limit on the density of the emitting region.
If we require that $t_{{\rm cool}}<7$\,days, and assume
$\left\langle Z \right\rangle \approx 1.7$ and $T\approx 10^{4}$\,K, we find that
$n\gtorder 7\times10^{7}$\,cm$^{-3}$.
Combining this limit on $n$ along with Equation~\ref{eq:n} and the fact that $\Delta{t}\ge20$\,day,
we get
\begin{equation}
\epsilon_{R}\ltorder 3.4\times10^{3}.
\label{eq:epsR1}
\end{equation}

Next, an order-of-magnitude estimate for the photon diffusion timescale
(e.g., Popov 1993; Padmanabhan 2001) is given by
\begin{eqnarray}
t_{{\rm diff}}\approx & \frac{9\kappa M_{{\rm CSM}}}{4\pi^{3} c r_{{\rm CSM}}} \cr
   \approx 0.32  & \Big(\frac{1}{\epsilon_{R}}\Big)
                 \Big(\frac{\kappa}{0.34\,{\rm cm\,gr}^{-1}}\Big)
                 \Big(\frac{\left\langle L \right\rangle}{1.7\times10^{40}\,{\rm erg\,s}^{-1}}\Big)  \cr
   \times      &  \Big(\frac{v_{{\rm CSM}}}{1000\,{\rm km\,s}^{-1}}\Big)^{-3}\,{\rm day}
\label{eq:tdiff}
\end{eqnarray}
Assuming that $t_{{\rm diff}}\ltorder 7$\,day, we can set the following
lower limit:
\begin{equation}
\epsilon_{R}\gtorder 0.04.
\label{eq:epsR2}
\end{equation}

Until now, our constraints on the radiative efficiency
are based on the properties of the precursor.
Next, we will use the properties of the May 2015 event to derive additional
constraints.

Assuming $v_{{\rm CSM}}\approx 1000$\,km\,s$^{-1}$,
after 500\,days (i.e., the time between the Dec. 2013 precursor and the May 2015 event),
the CSM traveled a distance of $r_{{\rm CSM}} \approx 4\times10^{15}$\,cm.
Since the rise time of the May 2015 event is about 2\,days (Table~\ref{tab:SNprop}),
we can use the diffusion timescale (Eq.~\ref{eq:tdiff}) to set an 
order-of-magnitude upper limit on the mass of the CSM:
%under the assumption that the transient is due to collision with the mass ejected during the progenitor
\begin{equation}
M_{{\rm CSM}}\ltorder 0.4 \Big(\frac{t_{{\rm diff}}}{2\,{\rm day}}\Big)
                       \Big(\frac{r_{{\rm CSM}}}{4\times10^{15}\,{\rm cm}}\Big) 
                       \Big(\frac{\kappa}{0.34\,{\rm cm\,g}^{-1}}\Big)^{-1}\,{\rm M}_{\odot}.
\label{eq:McsmL}
\end{equation}
Inserting this limit on $M_{{\rm CSM}}$ into Equation~\ref{eq:epsR} gives
\begin{equation}
\epsilon_{R}\gtorder 0.007 \Big(\frac{\Delta{t}}{20\,{\rm day}}\Big)
                          \Big(\frac{v_{{\rm CSM}}}{1000\,{\rm km\,s}^{-1}}\Big)^{-2}.
\label{eq:epsR3}
\end{equation}

We note that currently we also have the following upper limit on the
duration and luminosity of the precursor.
Since we did~not find a transient at the location of PTF\,13efv
in the Catalina Real Time Survey\footnote{http://nunuku.caltech.edu/cgi-bin/getcssconedb\_release\_img.cgi} (CRTS; Drake et al. 2009), and assuming CRTS can detect magnitude 19 transients, we can
conclude that the luminosity of the precursor was~not larger by more than factor
of about four than the observed luminosity of $1.7\times10^{40}$\,erg\,s$^{-1}$
seen during December 2013.
Furthermore, the PTF $g$-band nondetection prior to the December 2013 event
sets an upper limit of about 240\,days on $\Delta{t}$.
Since the precursor was~not detected
in the UV (Table~\ref{tab:SNprop}),
the bolometric correction is likely small.

To conclude, combining all the constraints,
we set the following limits on the radiative efficiency:
\begin{equation}
0.04 \ltorder \epsilon_{R}\ltorder 3400.
\label{eq:epsRf}
\end{equation}
We stress that this is an order-of-magnitude
estimate and it includes several assumptions
that are not necessarily correct.
Therefore, the results of this analysis should be viewed with caution.
Since we cannot determine whether the efficiency is smaller or larger than unity,
we cannot point definitively toward one of two types of scenarios:
kinetic energy converted into radiation or radiation-driven mass loss.
However, with improved observational constraints
this analysis can be used in the future to obtain
better estimates of the radiative efficiency of precursors.

\subsection{What Drives the Mass Loss and Radiation?}
\label{sec:What}

In the case that the 2013 event is caused by
a super-Eddington continuum-driven wind, we expect
it will satisfy a mass-loss vs.
luminosity relation.
In this case, Shaviv (2001) has shown that the total mass loss is given by
\begin{equation}
M \approx \mathcal{W} \frac{L-L_{{\rm Edd}}}{c\,c_{{\rm s}}}\Delta{t},
\label{eq:Shaviv}
\end{equation}
where $\mathcal{W}$ is a dimensionless constant that
empirically was found to be of order a few,
$c_{{\rm s}}$ is the speed of sound at the base of the wind
(estimated to be 60\,km\,s$^{-1}$),
and $L_{{\rm Edd}}$ is the Eddington luminosity. 
For $L \gg L_{{\rm Edd}}$ we can write
\begin{equation}
M \approx 4\times10^{-5} \Big(\frac{\mathcal{W}}{5}\Big) \Big(\frac{L}{1.7\times10^{40}\,{\rm erg\,s}^{-1}}\Big) \Big(\frac{\Delta{t}}{20\,{\rm day}}\Big)\,{\rm M}_{\odot}.
\label{eq:Shaviv1}
\end{equation}
This estimate is below the derived upper limits on the CSM mass,
and hence we cannot rule out this model.

The second option we would like to consider is that the radiation is
generated from conversion of the kinetic energy of the ejected mass
to radiation via interaction with previously emitted material (e.g., Smith et al. 2014).
One possible problem with this scenario is that the interaction
will produce mostly hard X-ray photons
(e.g., Fransson 1982; Katz et al. 2011; Murase et al. 2011, 2014; Chevalier \& Irwin 2012; Svirski et al. 2012).
Presumably, it is possible to convert these X-ray photons to visible light
by Comptonization or bound-free absorption (e.g., Chevalier \& Irwin 2012; Svirski et al. 2012).
Comptonization requires larger than unity Thompson optical depth,
while the bound-free absorption will need neutral CSM mass with
column densities above $\sim10^{23}$\,cm$^{-2}$.
We note that in the current
observations there is no evidence for a large Thompson optical depth,
but we cannot rule out strong bound-free absorption.
Moreover, this scenario may still work,
if we introduce large departures from
the spherical symmetry we have assumed so far.

\section{Summary}
\label{sec:Sum}

We present observations of a precursor, peaking at an absolute magnitude
of about $-12$, $\sim500$\,days prior to the SNHunt\,275 May 2015 event.
Also included are {\it Swift}-UVOT
observations of the May 2015 event that peaked at an absolute magnitude of $-17$.
We discuss the nature of the May 2015 event, and conclude that it is not yet
clear whether this event signals the final explosion of the progenitor or is
still another eruption.
If the latter, then we may
detect a SN taking place within months to a few years.

Finally, we use the observations to constrain the ratio
of the radiated energy to the kinetic energy of the precursor
(i.e., the radiative efficiency).
Under some simplistic assumptions, our order-of-magnitude estimate suggests
that the radiative efficiency of PTF\,13efv is $\gtorder 0.04$.
However, this still does not necessarily mean that all precursors
have similar radiative efficiencies.

\acknowledgments

We thank M. Graham, P. Kelly, A. Bostroem, I. Shivvers, and W. Zheng
for their help obtaining some of the optical spectra. We 
are also grateful to the staffs of the Palomar, Lick, and Keck Observatories
for their excellent assistance.
Research at Lick Observatory is partially supported by a generous gift
from Google.
Some of the data
presented herein were obtained at the W. M. Keck Observatory, which is
operated as a scientific partnership among the California Institute of
Technology, the University of California, and NASA; the observatory was
made possible by the generous financial support of the W. M. Keck
Foundation. The authors wish to recognize and acknowledge the very
significant cultural role and reverence that the summit of Mauna Kea
has always had within the indigenous Hawaiian community.  We are most
fortunate to have the opportunity to conduct observations from this
mountain.
E.O.O. is incumbent of
the Arye Dissentshik career development chair and
is grateful for support by
grants from the 
Willner Family Leadership Institute
Ilan Gluzman (Secaucus NJ),
Israel Science Foundation,
Minerva,
Weizmann-UK,
and the I-Core program by the Israeli Committee for Planning
and Budgeting and the Israel Science Foundation (ISF).
N.J.S. is grateful for the IBM Einstein Fellowship.
A.G.Y. is supported by the EU/FP7 via ERC grant no. 307260, the Quantum
Universe I-Core program by the Israeli Committee for Planning and
Budgeting and the ISF; by Minerva and ISF grants; by the Weizmann-UK
``making connections'' program; and by Kimmel and ARCHES awards.
M.S. acknowledges support from the Royal Society and EU/FP7-ERC grant 
no. 615929.
A.V.F.'s research is supported by the Christopher R. Redlich Fund, the 
TABASGO Foundation, and NSF grant AST-1211916.

\end{document}